\documentclass[aps,prl,amsmath,amssymb,footinbib,showpacs,twocolumn]{revtex4-1}
\usepackage{amsmath}
\usepackage{amssymb}
\usepackage{amsthm}
\usepackage{setspace}
\usepackage{graphicx}
\usepackage{braket}
\usepackage{mathrsfs}
\usepackage{float}
\usepackage[colorlinks = true,linkcolor = red,urlcolor  = blue,citecolor = blue,anchorcolor = blue]{hyperref}
\usepackage[utf8]{inputenc}
\usepackage[english]{babel}
\usepackage{bm}

\graphicspath{ {Figures/} }

\begin{document}

\title{Electrostatic effects and topological superconductivity in semiconductor-superconductor-magnetic insulator hybrid wires}

\author{Benjamin D. Woods}

\author{Tudor D. Stanescu}
\affiliation{Department of Physics and Astronomy, West Virginia University, Morgantown, WV 26506, USA}

\begin{abstract}
We investigate the impact of electrostatics on the proximity effect between a magnetic insulator and a semiconductor wire in semiconductor-superconductor-magnetic insulator hybrid structures. By performing self-consistent Schr$\ddot{\rm o}$dinger-Poisson calculations using an effective model of the hybrid system, we find that large effective Zeeman fields consistent with the emergence of topological superconductivity emerge within a large parameter window in wires with overlapping layers of magnetic insulator and superconductor, but not in non-overlapping structures. We show that this behavior is essentially the result of electrostatic effects controlling the amplitude of the low-energy wave functions near the semiconductor-magnetic insulator interface.
\end{abstract}
	
\maketitle

The successful experimental realization of Majorana zero modes (MZMs) --  non-Abelian anyons \cite{Stern2010} representing the condensed matter analogues of Majorana fermions \cite{Majorana1937,Kitaev2001} that provide a promising platform for topological quantum computing \cite{Kitaev2003,Nayak2008,DSarma2015,Stanescu2017} -- depends critically on the robustness of the topological superconducting phase that hosts them \cite{Alicea2012,Leijnse2012,Stanescu2013,Beenakker2013,Elliot2015,Aguado2017,Lutchyn2018}. In the absence of naturally occurring one-dimensional topological superconductors, the research has focused on hybrid structures \cite{Fu2008,Sau2010a,NadjPerge2013}, particularly semiconductor (SM) wires proximity coupled to s-wave superconductors (SCs) in the presence of a magnetic field parallel to the wire \cite{Lutchyn2010,Oreg2010,Mourik2012,Das2012,Finck2013,Albrecht2016,Nichele2017}. A large-enough field-induced Zeeman splitting ensures the emergence of a topological superconducting phase, even in the presence of some weak/moderate system inhomogeneity. However, in addition to suppressing the gap of the parent superconductor, in which orbital effects play an important role \cite{Nijholt2016} and which severely limits the realization of robust topological superconductivity, the applied magnetic field imposes serious constraints on the possible device layout for Majorana-based topological qubits \cite{Karzig2017}. 

A possible solution is to create the required Zeeman field by proximity coupling the semiconductor to a magnetic insulator \cite{Sau2010a,Sau2010}. Recently, this possibility has been explored experimentally using InAs nanowires with epitaxial layers of superconducting Al and ferromagnetic EuS \cite{Vaitiekenas2020,Liu2020a,Liu2020b}. 
A key finding was that an effective Zeeman field $\Gamma_{eff}^{SC}$ of order $1$ T ($\sim 0.05~$meV) emerges in the superconductor in the absence of an applied magnetic field, but only in nanowires with {\em overlapping} shells of superconductor and ferromagnetic insulator \cite{Vaitiekenas2020}.  Correlated with the emergence  of an effective Zeeman  field in the superconductor was the observation of zero-bias conductance peaks for charge tunneling into the end of the semiconductor wire,  which is consistent with the presence of topological superconductivity. These features are absent in  hybrid structures with {\em non-overlapping} Al and EuS covered facets \cite{Vaitiekenas2020}. 

The crucial question concerns the physical mechanism responsible for the startling contrast between the phenomenologies observed in the two setups. Furthermore, one may ask if, based on the understanding of this mechanism, one can identify efficient knobs for controlling the  magnitude of the effective Zeeman field emerging in the nanowire, to ensure that the topological superconducting phase is accessible and robust. 

A natural candidate for explaining the difference between the behaviors associated with the two setups is the ferromagnetic exchange coupling occurring inside the SC in the overlapping geometry due to spin-dependent scattering at the Al-EuS interface \cite{Tedrow1986,Tokuyasu1988,Bergeret2004,Strambini2017,Heikkila2019}. In turn, the proximity effect generated by the exchange-coupled superconductor inside the spin-orbit coupled nanowire could lead to the emergence of a topological superconducting state. In this scenario, the effective Zeeman  field $\Gamma_{eff}^{SM}$  required to drive the SM nanowire into the topological regime is induced ``indirectly'', via the Al layer. 
Consequently ,  it is controlled by the strength $\widetilde{\gamma}$ of the effective coupling between the semiconductor and superconductor, which also determines the size of the induced superconducting gap and the critical Zeeman field associated with the topological quantum phase transition (TQPT) \cite{Stanescu2017a}.  In particular, the minimum value of the critical field is given by $\widetilde{\gamma}$ and can be significantly larger that the induced gap in the strong coupling limit \cite{Stanescu2017a}. This poses a serious problem for the ``mediated proximity'' scenario. As explicitly shown below, the topological condition $\Gamma_{eff}^{SM} > \widetilde{\gamma}$ is inconsistent with the experimental parameters reported in Ref. \onlinecite{Vaitiekenas2020} and,  more importantly, is generally inconsistent with robust topological superconductivity. 

\begin{figure}[t]
\begin{center}
\includegraphics[width=0.45\textwidth]{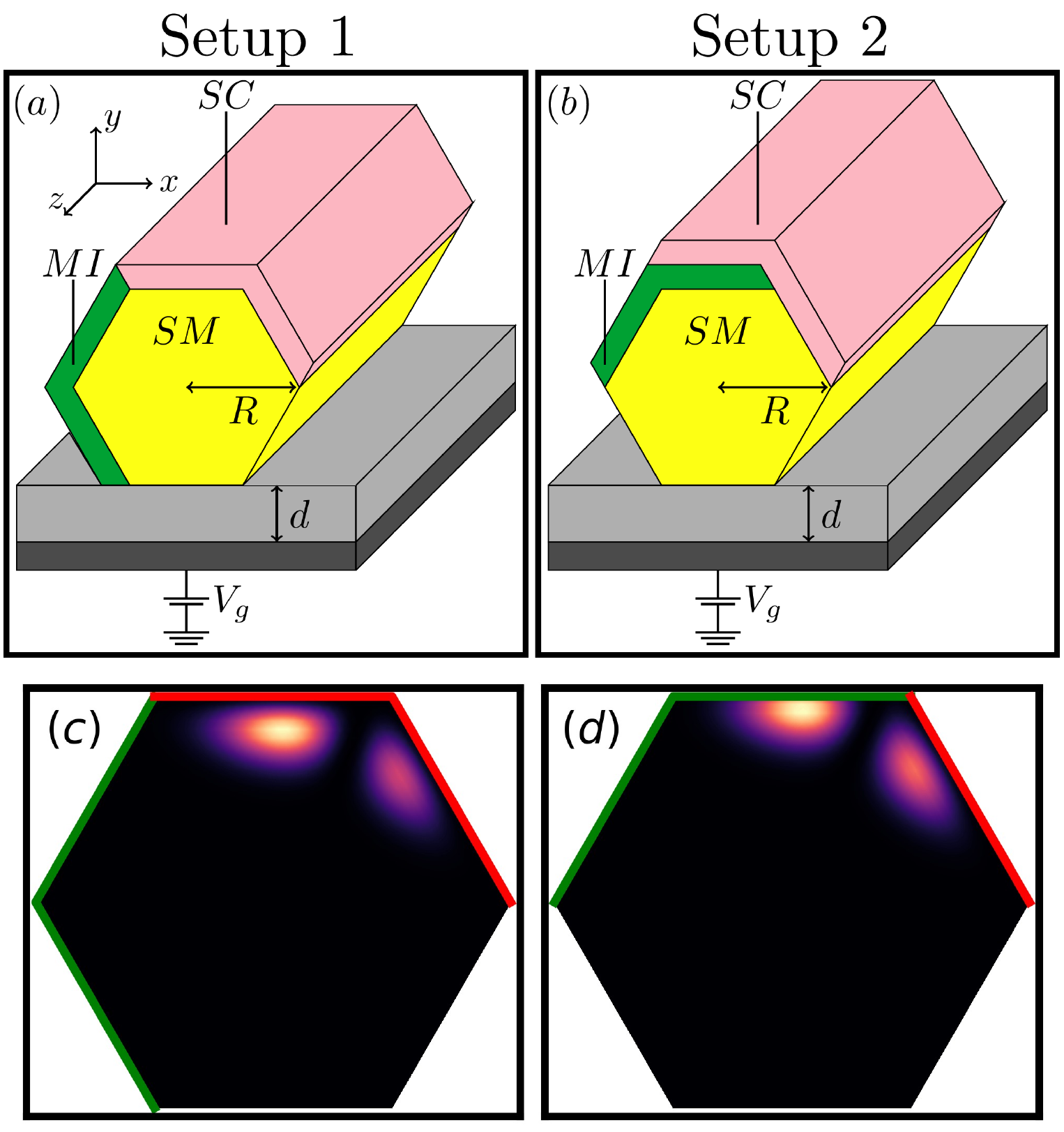}
\end{center}
\vspace{-2mm}
\caption{{\em Top panels}: Schematic representation of the hybrid structure studied in this work corresponding to a semiconductor nanowire (yellow) with (a) non-overlapping (setup 1) and (b) overlapping (setup 2)  layers of superconductor (pink) and magnetic insulator (green). An external potential is applied using a back-gate (black) separated from the wire by an insulating dielectric layer (gray). Parameters: $R = 50~$nm, $d = 10~$nm.
{\em Bottom panels}: Wave function profile of the second lowest transverse mode for parameters  corresponding to this mode being near the Fermi level: $V_g\approx-1.1~$V, $V_{SC}=0.15~$V.}
\label{FIG1}
\vspace{-2mm}
\end{figure}

In this paper we investigate a different scenario involving the ``direct'' proximity effect at the semiconductor - magnetic insulator (SM-MI) interface. We show that the strength of the effective Zeeman field $\Gamma_{eff}^{SM}$ induced in the wire by proximity to the MI is controlled by electrostatic effects, which, in turn, depend on the geometry of the SC layer and on the applied gate potential. In essence, because of the finite work function difference between the SM wire and the SC shell, the wave functions characterizing the low-energy states in the wire are strongly ``attracted'' toward the superconductor, regardless of whether the SM and SC are in direct contact or separated by a MI layer. This means away from the SM-MI interface in the non-overlapping setup and toward the SM-MI interface in the system with overlapping MI and SC layers (see Fig. \ref{FIG1}). As a result,  the induced $\Gamma_{eff}^{SM}$ has significantly higher values in the overlapping structure as compared to the non-overlapping setup. By performing self-consistent Schr$\ddot{\rm o}$dinger-Poisson calculations, we demonstrate that the overlapping setup is consistent with the emergence of topological superconductivity over a large window of system parameters and applied gate potentials, in sharp contrast with the non-overlapping structure.  Our findings support the feasibility of topological superconductivity in SM-SC-MI  hybrid structures and provide guidance for controlling the system and enhancing the robustness of the topological phase. 

Before we present our model calculations, let us briefly discuss the ``mediated proximity'' scenario. In the strong coupling limit, satisfying the topological  condition $\Gamma_{eff}^{SM} > \widetilde{\gamma}$ requires  a large effective Zeeman field $\Gamma_{eff}^{SC}$ inside the SC, possibly exceeding the Chandrasekhar-Clogston limit \cite{Chandrasekhar1962,Clogston1962}. Even assuming that spin-orbit coupling induced by proximity to the SM wire prevents the closing of the SC gap, its value (and, implicitly, the size of the topological gap) will be very small. On the other hand, in the weak/intermediate regime the induced SC gap and effective Zeeman field are approximately given by \cite{Stanescu2017a}
\begin{equation}
\Delta_{ind} \approx \frac{\widetilde{\gamma}~\!\Delta}{\widetilde{\gamma}+\Delta}, ~~~~~~~~~~~ \Gamma_{eff}^{SM} \approx \frac{\widetilde{\gamma}~\!\Gamma_{eff}^{SC}}{\widetilde{\gamma}+\Delta},   \label{Eq1}
\end{equation}
where $\Delta$ is the order parameter of the parent SC in the presence of the ferromagnetic exchange coupling generated by the MI. Using these relations, we can rewrite the topological condition in terms of the effective Zeeman field inside the SC and the induced gap   as 
\begin{equation}
\Gamma_{eff}^{SC} > \frac{\Delta^2}{\Delta - \Delta_{ind}} \geq 4\Delta_{ind}. \label{Eq2}
\end{equation}
First, we notice that the parameters characterizing the recent experiment \cite{Vaitiekenas2020}, i.e., $\Gamma_{eff}^{SC}\sim \Delta_{ind}\sim 0.05~$meV,  do not satisfy Eq. (\ref{Eq2}). Second, we point out that satisfying Eq. (\ref{Eq2}) implies an induced gap $\Delta_{ind}$ significantly smaller than $0.18\Delta_0$, with $\Delta_0$ being the bare SC gap,  a  value that would require an effective Zeeman field $\Gamma_{eff}^{SC}$ comparable to the  Chandrasekhar-Clogston limit and a sizable effective gap $\Delta > \Delta_{ind}$. This  is set of conditions that would be extremely hard to satisfy, even assuming the presence of proximity-induced spin-orbit coupling in the parent SC. For example, an effective Zeeman field  $\Gamma_{eff}^{SC} \sim 0.05~$meV (similar to the experimental value) can only generate topological superconductivity with a gap smaller than $12.5~\mu$eV. These considerations lead to the conclusion that the ``mediated proximity'' mechanism does not enable the realization of robust topological superconductivity in SM-SC-MI hybrid structures and suggest that the investigation of the  ``direct'' proximity effect at the SM-MI interface is critical for understanding the low-energy physics in these systems. 

The SM-SC-MI hybrid system studied in this work is represented schematically in the top panels of Fig. \ref{FIG1}. We focus on two setups corresponding to the non-overlapping (setup 1 in Fig.  \ref{FIG1}) and overlapping (setup 2) configurations investigated in the recent experiment \cite{Vaitiekenas2020}. We do not address explicitly the proximity effect between the MI and the SC (in setup 2), but focus instead on the impact of electrostatics on the proximity-induced Zeeman field and pairing potential at the SM-MI and SM-SC interfaces, respectively. After integrating out the MI and SC degrees of freedom, the SM wire is described by the energy-dependent effective ``Hamiltonian,'' 
\begin{equation}
    H_{eff}\left(k_z,\omega\right) = 
    H_{SM}(k_z) + \Sigma_{MI}\left(k_z,\omega\right) 
    + \Sigma_{SC}\left(k_z,\omega\right), \label{Heff}
\end{equation}
where $k_z$ is the wave number along the wire axis, $H_{SM}$ is the bare SM Hamiltonian, and $\Sigma_{MI}$ and $\Sigma_{SC}$ are self-energies characterizing the proximity effects at the SM-MI and SM-SC interfaces, respectively. The key parameter characterizing $\Sigma_{MI}$ is the bare Zeeman field $\Gamma$ of the magnetic insulator, which we treat as a free phenomenological parameter. To control the SM-SC coupling strength, we include a tunable potential  barrier at the SM-SC interface. Hence, the key parameters associated with $\Sigma_{SC}$ are the parent SC pairing potential $\Delta_o$ and the SM-SC interface barrier strength $V_{b}$.
The bare SM Hamiltonian has the form 
\begin{equation}
H_{SM} =\left[ \frac{\hbar^2}{2m^*} \left(-\bm{\nabla}_\perp^2 + k_z^2\right) - \mu_{SM} - e\phi\left({\bm r}_\perp\right) + \alpha k_z\sigma_y \right]\tau_z, \label{HSM} 
\end{equation}
where $m^*$ is the effective mass, $\bm \nabla_\perp$ is the nabla operator acting in the $xy\!-\!\text{plane}$, $\mu_{SM}$ is the chemical potential of the SM, $\phi$ is the electrostatic potential, $\alpha$ is the Rashba spin-orbit coefficient, and $\sigma_i$ and $\tau_i$ are Pauli matrices acting in the spin and particle-hole spaces, respectively. The electrostatic potential satisfies the Poisson equation, 
\begin{equation}
    -{\bm \nabla}_\perp \cdot \varepsilon({\bm r}_\perp){\bm \nabla}_\perp\phi ({\bm r}_\perp)= \rho({\bm r}_\perp), \label{Poisson}
\end{equation}
where $\varepsilon$ is the region-dependent dielectric constant inside the SM, insulating layer, and vacuum and $\rho$ is the charge density inside the wire. The potential satisfies Dirichlet boundary conditions on the back-gate, $\phi=V_g$, and superconductor, $\phi=V_{SC}$. The charge density $\rho$  is determined by the occupied states of the Hamiltonian (\ref{HSM}). Equations (\ref{HSM}) and (\ref{Poisson}) form a set of  coupled Schr\"{o}dinger-Poisson equations that are solved self-consistently. Details regarding the model and the numerical procedure are provided in the Supplementary Material. 

\begin{figure}[t]
\begin{center}
\includegraphics[width=0.45\textwidth]{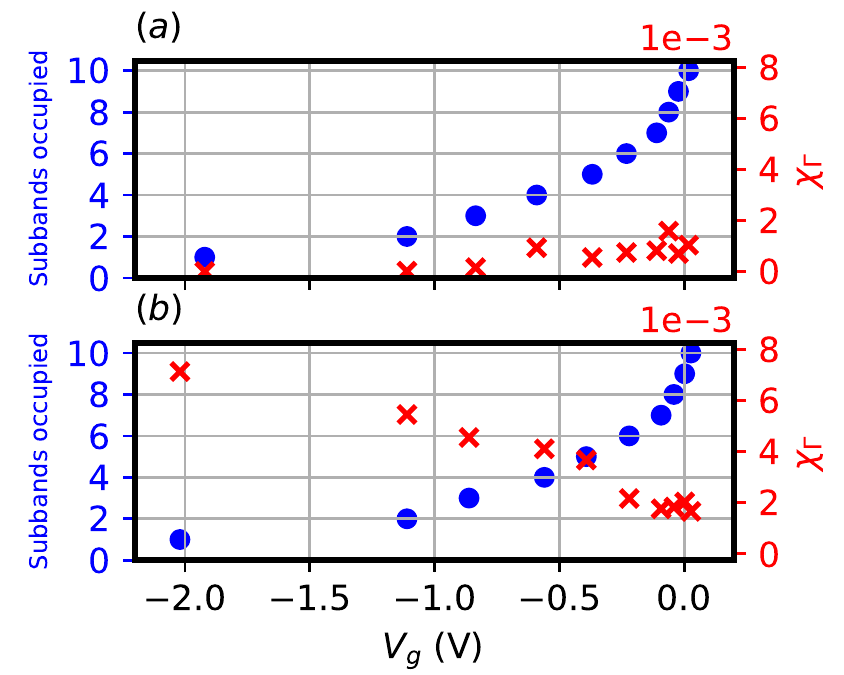}
\end{center}
\vspace{-7mm}
\caption{{\em Blue disks}: Subband occupancy as a function of the applied gate potential $V_g$ for (a) setup 1  and (b) setup 2. The values of $V_g$ correspond to the bottom of a certain subband $n$ being at the chemical potential. Note the negligible difference between the two setups. {\em Red crosses}: Linear susceptibility $\chi_\Gamma = \Gamma_{eff}^{SM}/\Gamma$ characterizing the direct proximity effect at the SM-MI interface. Note the significant difference between (a) setup 1 and (b) setup 2, indicative of a much stronger direct SM-MI proximity effect in the overlapping configuration as compared to the non-overlapping setup. 
The system parameters are: 
$V_{b} \rightarrow \infty$, $\Gamma \rightarrow 0$, and $V_{SC} = 0.15~$V.}
\label{FIG2}
\vspace{-2mm}
\end{figure}

First, we determine the dependence of the number of occupied subbands on the applied gate potential ($V_g$) and identify the values of $V_g$ corresponding to the bottom of a certain subband $n$ being at the chemical potential, which provides the optimal condition for the emergence of topological superconductivity. The results are shown in Fig. \ref{FIG2} (blue disks). Note that the differences in subband occupancy between setups 1 and 2 are very small, which demonstrates that electrostatic effects depend weakly on the location of the magnetic insulator layer. Next, we solve the equation $H_{eff} \psi = \omega\psi$ corresponding to $V_{b} \rightarrow \infty$ and $\Gamma \rightarrow 0$ and calculate the (linear) susceptibility $\chi_\Gamma = \Gamma_{eff}^{SM}/\Gamma$ characterizing the direct proximity effect at the SM-MI interface. The results are shown in Fig. \ref{FIG2} (red crosses). Note the striking difference between the two setups. Particularly significant is that in the low-occupancy regime $n\lesssim 5$, which is expected to be most favorable for realizing robust topological superconductivity \cite{Woods2020}, the susceptibility for setup 2 (overlapping layers) is 5-50 times higher than the corresponding susceptibility for setup 1. This behavior is determined by electrostatic effects, which result in the wave function of the lowest energy mode (relative to the Fermi level) being localized in the vicinity of the superconductor, as shown, for example, in Fig. \ref{FIG1} (c) and (d). For setup 1, this implies a wave function localized away from the SM-MI interface (hence, weak SM-MI proximity effect), while for setup 2 the wave function has a significant amplitude at the interface with the magnetic insulator, leading to large values of $\chi_\Gamma$. We note that the wave functions associated with higher energy transverse modes are more delocalized, reducing the difference between the two setups. However, the high-occupancy regime is characterized by a small inter-subband spacing, which makes the topological phase susceptible to disorder and other types of system inhomogeneity \cite{Woods2020}.

\begin{figure}[t]
\begin{center}
\includegraphics[width=0.45\textwidth]{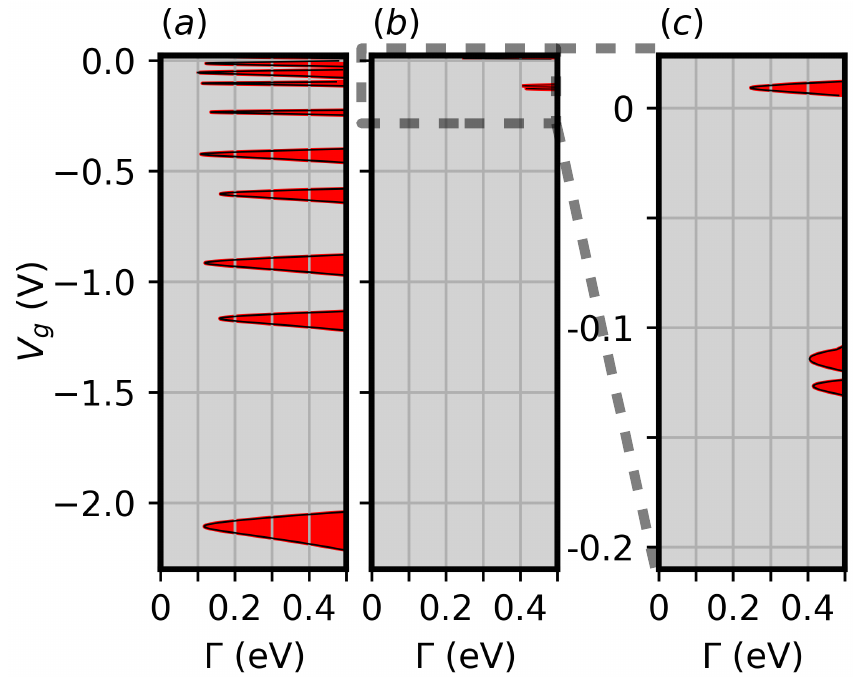}
\end{center}
\vspace{-5mm}
\caption{Topological phase diagram as function of the (phenomenological) Zeeman field $\Gamma$ characterizing the MI and the applied gate potential $V_g$ for (a) the overlapping structure (setup 2) and (b) the non-overlapping structure (setup 1). Panel (c) is an inset corresponding to the high occupancy regime in panel (b). Note that setup 2 is consistent with the emergence of topological superconductivity within a significant parameter window, in sharp contrast with setup 1. The system parameters are: $V_{SC} = 0.15~$V and $V_{b} = 2.75$~$\text{eV} \cdot \text{nm}$}
\label{FIG3}
\vspace{-2mm}
\end{figure}

Having elucidated the key role played by electrostatics in determining the strength of the SM-MI proximity effect, we calculate the topological phase diagram as a function of the bare Zeeman field $\Gamma$ and the applied gate voltage $V_g$ for a hybrid system with $\Delta_o= 0.3~$meV \cite{Court2007} and a SM-SC interface barrier $V_{b} = 2.75~\text{eV}\cdot\text{nm}$. The phase boundary separating the trivial and topological superconducting phases are obtained by finding $\Gamma$ such that the equation $H_{eff}(k_z=0, \omega=0) \psi =0$ is satisfied. The results are shown in Fig. \ref{FIG3}. Note that the overlapping configuration (setup 2) is consistent with the emergence of a topological phase for $\Gamma\sim 100-200~$meV and $V_g$ near the optimal values corresponding to the bottom of a certain subband being at the Fermi level [see Fig. \ref{FIG3}(a)]. By contrast, the non-overlapping structure (setup 1) cannot support topological superconductivity for $\Gamma < 500~$meV, except in the high-occupancy regime [see Fig. \ref{FIG3}(b) and (c)].  We emphasize that including the ``indirect'' proximity effect for setup 2, reduces the parent SC gap $\Delta$ and generates an effective Zeeman field $\Gamma_{eff}^{SC}$ inside the parent superconductor, which favors the emergence of topological superconductivity and further enhances the already substantial difference between the two setups.  

\begin{figure}[t]
\begin{center}
\includegraphics[width=0.44\textwidth]{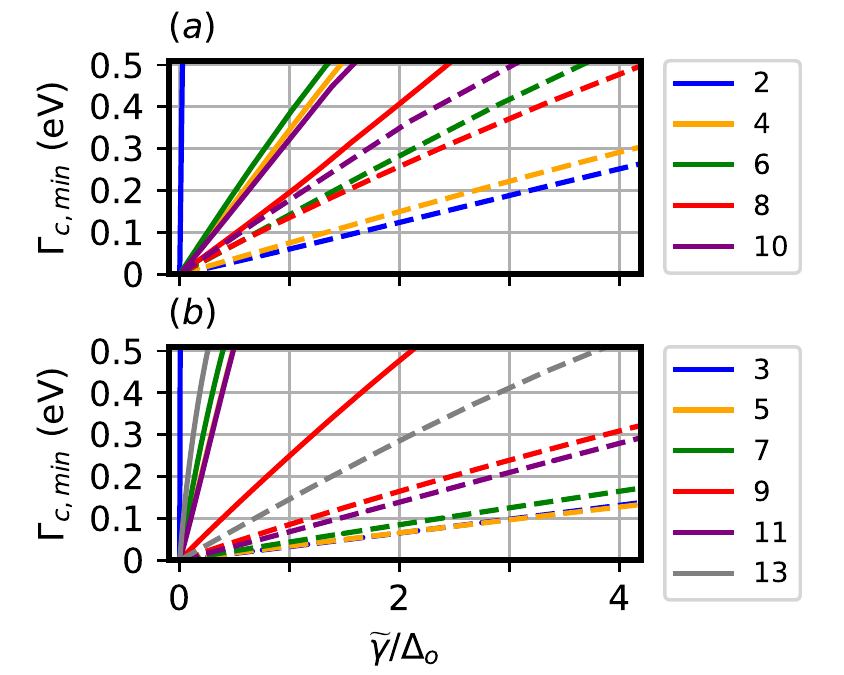}
\end{center}
\vspace{-5mm}
\caption{Dependence of the minimum critical Zeeman field $\Gamma_{c, min}^n$  on the effective SM-SC coupling $\widetilde{\gamma}$ for two values of the SM-SC work function difference:  (a) $V_{SC} = 0.15~$V and (b) $V_{SC} = 0.3~$V.  The full lines correspond to setup 1, while the dashed lines are for the overlapping structure (setup 2). Note that the minimum critical Zeeman fields required for the emergence of a topological SC phase are systematically larger (by up to three orders of magnitude) in the non-overlapping configuration as compared to the overlapping setup.}
\label{FIG4}
\vspace{-2mm}
\end{figure}

To demonstrate the robustness of our results, we investigate the dependence of the minimum critical field $\Gamma_{c, min}^n$ characterizing the topological phase transition associated with  subband  $n$ on the strength of the effective SM-SC coupling $\widetilde{\gamma}$ for two values of the SM-SC work function difference $V_{SC}$. The effective coupling is calculated from Eq. (\ref{Eq1}) as $\widetilde{\gamma} = \Delta_{ind}\sqrt{\Delta_o + \Delta_{ind}}/\sqrt{\Delta_o - \Delta_{ind}}$ \cite{Stanescu2017a}, where $\Delta_{ind}$ is the induced gap for $\Gamma_{eff}^{SM} = 0$.  The results shown in Fig. \ref{FIG4} confirm the striking difference between the overlapping (dashed lines) and the non-overlapping (solid lines) setups. More specifically, the (bare) minimum critical Zeeman fields required for the emergence of a topological SC phase are systematically larger (by up to three orders of magnitude) in the non-overlapping configurations as compared to the overlapping setup. 
A comparison between panels (a) and (b) in Fig. \ref{FIG4} shows that this trend increases with $V_{SC}$. Including the ``indirect'' proximity effect can only enhance the difference between the two configurations. 

In conclusion, we have demonstrated that electrostatic effects play a critical role in determining the strength of the (direct) proximity effect between a magnetic insulator and a semiconductor wire in semiconductor-superconductor-magnetic insulator (SM-SC-MI) hybrid structures. These electrostatic effects are controlled by the applied gate potential  and by the geometry of the superconducting layer, regardless of whether the SM and SC are in direct contact or separated by a MI layer. We have argued that the ``indirect'' proximity effect emerging in structures with overlapping SC and MI layers is generally insufficient for the realization of a  topological superconducting phase in the hybrid system. However, in these overlapping structures electrostatics favors the realization of low-energy transverse modes with large amplitudes near the SM-MI interface, which, in turn, results in a strong proximity effect between the MI and the SM wire and the emergence of a large effective Zeeman field consistent with the presence topological superconductivity. By contrast, such large proximity-induced Zeeman fields do not occur in non-overlapping structures within similar parameter windows. On the one hand, our results suggest that the recently reported experimental findings \cite{Vaitiekenas2020} are consistent with the presence of small proximity-induced Zeeman fields and topologically-trivial superconductivity in non-overlapping structures and significant effective Zeeman fields in the overlapping setup, large-enough to generate topological superconductivity in a homogeneous system. On the other hand, our findings suggest possible strategies for enhancing the robustness of the topological superconducting phase realized in a SM-SC-MI hybrid system. For example, using a lateral gate (instead of or in addition to a back gate) may enable a better control of the amplitudes of the relevant wave functions at the interfaces between the SM wire and the magnetic insulator or the parent superconductor. In addition, changing the areas of the SM-SC and SC-MI interfaces (e.g., having three facets covered by superconductor) can significantly affect the strength of the induced SC pairing potential and effective Zeeman field. Finally, since the ``indirect'' proximity effect alone cannot generate topological superconductivity and is not required to generate it, as shown in this study,  but has the rather undesired effect of reducing the superconducting order parameter of the parent SC, it may be convenient to reduce the effective coupling at the SC-MI interface, e.g., by adding a thin nonmagnetic insulating layer. This would have a minimal impact on the electrostatics, while enhancing the induced SC gap. Of course, quantitative estimates of the topological gap within these scenarios require a more detailed modeling of the hybrid structure that explicitly includes the proximity effect at the SC-MI interface. 

%


\pagebreak
\widetext
\begin{center}
\textbf{\large Supplementary Material: Electrostatic effects and topological superconductivity in semiconductor-superconductor-magnetic insulator hybrid wires}
\end{center}
\setcounter{equation}{0}
\setcounter{figure}{0}
\setcounter{table}{0}
\makeatletter
\renewcommand{\theequation}{S\arabic{equation}}
\renewcommand{\thefigure}{S\arabic{figure}}
\renewcommand{\bibnumfmt}[1]{[S#1]}
\renewcommand{\citenumfont}[1]{S#1}

\section{Details of the model}
In this section, we provide details regarding the model used in the main text. The SM-SC-MI hybrid system, assumed to be infinitely long, is described by the Hamiltonian 
\begin{equation}
    H\left(k_z\right) = H_o\left(k_z\right) + H_{b}. \label{HAM}
\end{equation}
The first term, which includes the SM wire and the SC and MI layers, is given by
\begin{equation}
\begin{split}
    H_o\left(k_z\right) =&
    \bigg[
    -\bm\nabla_\perp \cdot \frac{\hbar^2}{2m^*\left(\bold{r}_\perp\right)} \bm\nabla_\perp 
    + \frac{\hbar^2k_z^2}{2m^*\left({\bm r}_\perp\right)}
    -\mu\left({\bm r}_\perp\right)
    - e\phi\left({\bm r}_\perp\right) + \Gamma\left({\bm r}_\perp\right)\sigma_z
    + \alpha\left({\bm r}_\perp\right)k_z\sigma_y \bigg]\tau_z
    + \Delta\left({\bm r}_\perp\right) \sigma_y \tau_y
\end{split}, \label{Ho} 
\end{equation}
where $m^*$ is the subsystem-dependent effective mass, $\bm \nabla_\perp$ is the nabla operator in the $xy\!-\!\text{plane}$ (i.e., transverse to the wire axis), $\mu$ is the chemical potential, $\phi$ is the electrostatic potential, $\Gamma$ is the Zeeman energy, $\alpha$ is the Rashba spin-orbit coefficient, $\Delta$ is the superconducting pairing, and $\sigma_i$ and $\tau_i$ are Pauli matrices acting in spin and particle-hole space, respectively. Note that these parameters are piece-wise functions with respect to the SM, SC, and MI regions. In particular, $\alpha$,  $\Delta$, and $\Gamma$ are uniform and non-zero only within the SM, SC, and MI regions, respectively. Each region, therefore, provides a necessary ingredient for topological superconductivity, as captured by, e.g., the minimal 1D models \cite{Oreg2010,Lutchyn2010}. For simplicity, we have neglected transverse spin-orbit coupling. The three regions are characterized by different effective masses and chemical potential values (relative to the bottom of the corresponding bands). Specifically,  we have $m^*_{SM} = 0.023 m_o$, $m^*_{SC} = m_o$, $m^*_{MI} =  m_o$ \cite{Chen2010} (where $m_o$ is the free electron mass), $\mu_{SM} = 0$, $\mu_{SC} = 10~$eV, and $\mu_{MI}=-1~$eV \cite{Liu2020a}. We model the SC region of the device as a semi-infinite bulk superconductor, which avoids including disorder as an ingredient needed to reproduce the experimentally observed induced gaps \cite{Antipov2018} for systems with thin superconductor layers.  A bulk superconducting region is attached to each facet of the SM which is in direct contact with the SC. Note that we do not attach a bulk superconducting region to the top of the MI region in setup 2 (see Fig. 
\ref{FIG1}(b) in the main text), since we are not considering the direct proximity effects between the MI and SC regions. To control the coupling between the SM and bulk SC(s), we include a  barrier potential at the SM-SC interface, 
\begin{equation}
H_{b} = V_{b}
    \sum_n \delta\left(u_{n} - \sqrt{3}R/2\right)\tau_z, \label{Hbar}
\end{equation}
where $V_{b}$ is the barrier strength, the sum runs over all SM-SC interfaces, and $u_n$ is the coordinate normal to the $n^\text{th}$ SM-SC interface.

As stated in the main text, we integrate out the SC and MI degrees of freedom and obtain an energy-dependent effective ``Hamiltonian'' given by 
\begin{equation}
    H_{eff}\left(k_z,\omega\right) = 
    H_{SM}(k_z) + \Sigma_{MI}\left(k_z,\omega\right) 
    + \Sigma_{SC}\left(k_z,\omega\right), \label{Heff2}
\end{equation}
where $H_{SM}$ is the component of (\ref{HAM}) restricted to the SM region and the SM-MI and SM-SC interfaces, and $\Sigma_{MI}$ and $\Sigma_{SC}$ are the self-energies from the MI and SC, respectively, which are localized on the SM-MI and SM-SC interface degrees of freedom. We numerically calculate these self-energies using an accelerated iterative algorithm \cite{Sancho1985,Klimes2007} that allows for SC regions of thousands of meters in thickness. 

As stated in the main text, the electrostatic potential in (\ref{HAM}) is governed by the Poisson equation,
\begin{equation}
   -{\bm \nabla}_\perp \cdot \varepsilon({\bm r}_\perp){\bm \nabla}_\perp\phi ({\bm r}_\perp)= \rho({\bm r}_\perp), \label{Poisson2}
\end{equation}
where $\varepsilon$ is the region-dependent permittivity and $\rho$ is the charge density. The potential is subject to Dirichlet boundary conditions on the back-gate, $V_g$, and superconductor, $V_{SC}$, e.g. the superconductor is treated as a perfect metal with regards to the electrostatics. In addition, we impose von Neumann type boundary conditions on the top, left, and right surfaces of a box of side length $b$ surrounding the wire. We emphasize that the potential within the nanowire is negligibly affected by these boundary conditions – e.g., the exact value of $b$ or whether we choose von Neumann or Dirichlet boundary conditions on the top, left, and right surfaces of the box – provided $b \gg R$. The charge density in (\ref{Poisson2}) is determined by the occupied states of the Hamiltonian (\ref{HAM}), which in turn depend upon the electrostatic potential, $\phi$. Therefore, equations (\ref{HAM}, \ref{Poisson2}) represent a coupled set of Schr\"{o}dinger-Poisson equations requiring a self-consistent solution for the charge density and electrostatic potential. Solving the Schr\"{o}dinger-Poisson equations in the presence of the bulk superconductor is a non-trivial task due to the continuum of states with energies outside of the superconducting gap. We therefore work in the limit of $V_{b} \rightarrow \infty$ (e.g. the SC is uncoupled from the SM and MI regions within the Hamiltonian) when solving for the electrostatic potential, $\phi$, self-consistently. The resulting potential is then used in the Hamiltonian for non-infinite $V_{b}$. Note that the presence of the superconductor still plays a key role in determining the potential since it provides a boundary condition for the Poisson equation. Moreover, we set $\Gamma = 0$ and neglect the dynamical dependence of the MI self-energy, making the effective Hamiltonian energy independent, when solving for $\phi$ self-consistently. This static approximation is justified since the large negative value of $\mu_{MI}$ causes $\Sigma_{MI}$ to only weakly depend on $\omega$ over the energy range relevant to the Schr\"{o}dinger-Poisson problem. A simple iterative mixing scheme is used to solve the Schr\"{o}dinger-Poisson equations self-consistently with these approximations. The relative permittivity of the SM, MI, dielectric, and surrounding air are given by 15.2, 11.1, 24, and 1, respectively.

The eigenstates of the Hamiltonian (\ref{HAM}) are found using finite element methods \cite{Ramdas2007}. A linear Lagrange basis set is used on a structured triangular mesh with 2 nm between vertices within the SM region. The mesh within the MI and SC regions has a much smaller vertex spacing of 0.01 nm in the direction normal to the SM-MI and SM-SC interfaces to handle the large effective masses of these regions. The finite element package FEniCS \cite{Alnaes2015} is used when solving the Poisson equation (\ref{Poisson2}). Meshes for the Poisson equation are generated using Gmsh \cite{Geuzaine2009}. The characteristic element mesh sizes within the SM, MI, dielectric, and surrounding air are taken to be 1.5, 1, 2, and 10 nm, respectively.

\begin{figure}[t]
\begin{center}
\includegraphics[width=0.45\textwidth]{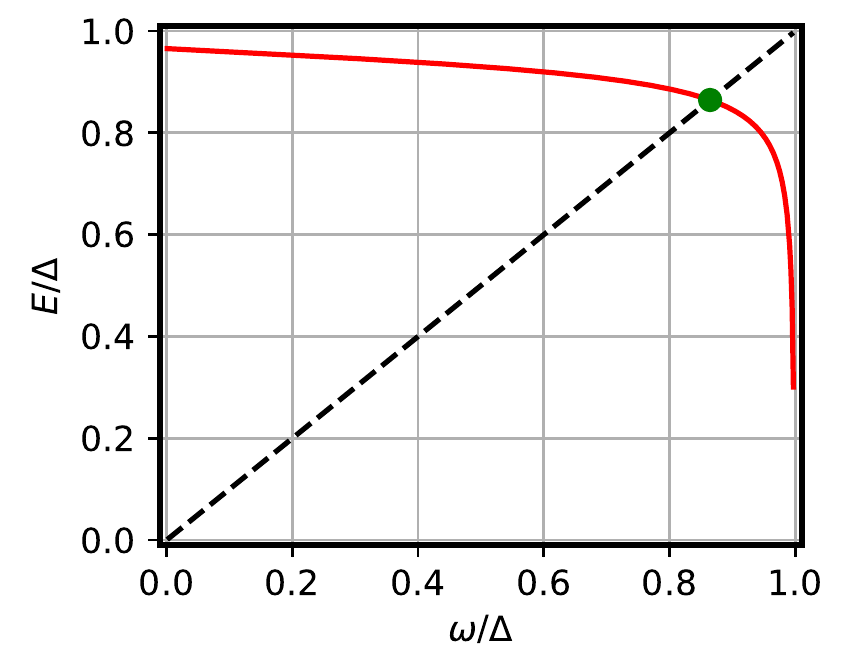}
\end{center}
\vspace{-5mm}
\caption{The lowest postive eigenvaule (red solid line) of $H_{eff}\left(k_z,\omega\right)$ is plotted as a function of input $\omega$ for some example system parameters with $\Gamma = 0$. The energy of a subgap state is found where the eigenvalue curve intersects (indicated by green dot) the line $E = \omega$ (black dashed line).}
\label{FIGS1}
\end{figure}

\section{Solving for subgap states}
In this section, we describe how we numerically solve for subgap states, e.g. states with energies withing the superconducting gap of the parent superconductor.
Both of the self-energies in (\ref{Heff2}) are purely real within the superconducting gap of the parent superconductor, implying that any subgap states must satisfy the eigenvalue equation, 
\begin{equation}
	H_{eff}\left(k_z,\omega\right) \psi\left(k_z\right) = \omega \psi\left(k_z\right), \label{SGE}
\end{equation}
where $|\omega| < |\Delta|$. Note that in (\ref{SGE}), $\omega$ appears both within the effective Hamiltonian and as the eigenvalue. Therefore, we must solve the eigenvalue equation self-consistently, e.g. the input $\omega$ needs to be equal to one of the eigenvalues of the effective Hamiltonian. To understand how we can find such as $\omega$, we plot in Fig. \ref{FIGS1} the lowest positive eigenvalue (red solid line) of $H_{eff}\left(k_z,\omega\right)$ as a function of $\omega$ for some example system parameters with $\Gamma = 0$. Equation (\ref{SGE}) is satisfied when the eigenvalue curve interests the line $E = \omega$ (black dashed line). We notice that the lowest positive eigenvalue of $H_{eff}\left(k_z,\omega\right)$ is monotonic over the range $-|\Delta| < \omega < |\Delta|$. Therefore, a simply bisection algorithm allows us to find $\omega$ satisfying (\ref{SGE}), provided it exists, by iteratively reducing the sub-interval in which $\omega = E_1(\omega)$ is possible, where $E_1$ is the lowest positive eigenvalue of the effective Hamiltonian.

\end{document}